\documentclass[%
 reprint,
superscriptaddress,
 amsmath,amssymb,
 aps,
longbibliography,
]{revtex4-1}

\usepackage{graphicx}
\usepackage{color}
\usepackage{dcolumn}
\usepackage{bm}
\usepackage[normalem]{ulem}

\begin{document}

\preprint{APS/123-QED}



\title{A CMOS dynamic random access architecture for radio-frequency readout of quantum devices.}

\author{S. Schaal}
\email{simon.schaal.15@ucl.ac.uk}
\affiliation{London Centre for Nanotechnology, University College London, London WC1H 0AH, United Kingdom}
\author{A. Rossi}
\affiliation{Cavendish Laboratory, University of Cambridge, J. J. Thomson Ave., Cambridge, CB3 0HE, United Kingdom}
\author{S. Barraud}
\affiliation{CEA, LETI, Minatec Campus, F-38054 Grenoble, France}
\author{J. J. L. Morton}
\affiliation{London Centre for Nanotechnology, University College London, London WC1H 0AH, United Kingdom}
\affiliation
{Department of Electronic \& Electrical Engineering, University College London, London WC1E 7JE, United Kingdom}
\author{M. F. Gonzalez-Zalba}
\email{mg507@cam.ac.uk}
\affiliation
{Hitachi Cambridge Laboratory, J.J. Thomson Avenue, Cambridge CB3 0HE, United Kingdom}

\date{\today}

\begin{abstract}
Quantum computing technology is maturing at a relentless pace, yet individual quantum bits are wired one by one. As quantum processors become more complex, they require efficient interfaces to deliver signals for control and readout while keeping the number of inputs manageable. Digital electronics offers solutions to the scaling challenge by leveraging established industrial infrastructure and applying it to integrate silicon-based quantum devices with conventional CMOS circuits. Here, we combine both technologies at milikelvin temperatures and demonstrate the building blocks of a dynamic random access architecture for efficient readout of complex quantum circuits.
Our circuit is divided into cells, each containing a CMOS quantum dot (QD) and a field-effect transistor that enables selective readout of the QD, as well as charge storage on the QD gate similar to 1T-1C DRAM technology. We show dynamic readout of two cells by interfacing them with a single radio-frequency resonator. Our results demonstrate a path to reducing the number of input lines per qubit and enable addressing of large-scale device arrays.
\end{abstract}

\pacs{Valid PACS appear here}
\maketitle



Quantum computers promise to solve problems which seem intractable using conventional computers~\cite{Montanaro2016QuantumOverview}. Several different physical implementations of a quantum computer are being developed~\cite{Ladd2010} and state-of-the-art processors are approaching the level of 50 to 100 quantum bits (or qubits)  where quantum computers are expected to demonstrate capabilities beyond conventional computers for specific tasks~\cite{Neill2018AQubits}. 

For most physical realizations, quantum processors require cryogenic temperatures to operate, precise low-noise control signals~\cite{VanDijk2018} to manipulate the information and highly sensitive readout techniques to extract the results; all without disturbing the fragile quantum states. In current solid-state quantum processors, signals are generated using general-purpose instruments at room temperature and delivered to the quantum processor at low temperatures. The physical qubits across all platforms are controlled directly with at least one control line per qubit. However, as the size of quantum processors continues to increase, the one-qubit-one-input approach is recognized to be unsustainable~\cite{Franke2018RentsComputing}, especially if we consider that a large-scale fault-tolerant quantum computer might ultimately require $10^8$ qubits to solve the most computationally demanding algorithms~\cite{Fowler2012}. Efficiently delivering control and readout signals to increasingly more complex quantum circuits while reducing the number of inputs per qubit is one of the major challenges towards a fully-fledged large-scale universal quantum computer.

Digital electronics may offer some opportunities to overcome this difficulty. Some of the challenges that large-scale quantum computing faces resemble those that conventional computing has already solved. A perfect example is controlling billions of transistors with just a few thousands of input-output connections. Moreover, integrated digital electronics allows signal generation, data flow management, low-level feedback and high-level operations locally. Therefore, to relax wiring requirements and reduce the latency of solid-state quantum computers, integration of digital electronics with quantum devices at cryogenic temperatures could be a promising strategy~\cite{Reilly2015,Hornibrook2015}. Nevertheless, to apply this approach, understanding the behaviour of digital circuits at cryogenic temperatures is paramount. 

Digital information processing devices are typically manufactured using silicon as the base material. Coincidentally, electron spins in silicon are amongst the most promising candidates for large-scale quantum computing~\cite{Kane1998,Loss1998} due to their small footprint (sub $100$~nm dimensions) and very long coherence times, particularly in isotopically purified silicon ($^{28}$Si)~\cite{Veldhorst2015,Muhonen2014}. Silicon-based spin qubits benefit from a variety of qubit designs and different coupling strategies~\cite{Veldhorst2015,Eng2015,Urdampilleta2015,Zajac2016,Mi2017,Hutin2016,House2015,Shi2012,Harvey-Collard2017,Kawakami2014,Watson2018,Zajac2017,Weber2014} and can be read out dispersively using Pauli spin-blockade~\cite{Betz2015,West2018Gate-basedSilicon,Pakkiam2018Single-shotSilicon}. To date, operation of one-dimensional arrays has been shown~\cite{Zajac2016}, high-fidelity single qubit gates~\cite{Veldhorst2014,Takeda2016,Yoneda2017,Huang2018} and two qubit gates~\cite{Veldhorst2015,Zajac2017} have been achieved and a programmable two-qubit silicon-based processor has been demonstrated~\cite{Watson2018}.


Recently, it was shown that CMOS transistors can be used as the basis for spin qubits~\cite{Hutin2016,Crippa2018}. Several other silicon-based quantum devices could, in principle, be realised in a manner compatible with industrial CMOS processes, with the potential of large-scale, high-yield fabrication. It seems then natural to explore the direct integration of silicon quantum devices and conventional CMOS digital technology to tackle the challenges in addressing, controlling and reading multi-qubit circuits.  Blueprints of such all-silicon systems integrating quantum and classical components have emerged~\cite{Veldhorst2017,Vandersypen2017,Li} as well as first basic demonstrations of direct integration~\cite{Schaal2018}.

%


Here, we present a CMOS dynamic random access architecture for control and readout of multiple quantum devices. Our design is inspired by the square arrays found in dynamic random access memory (1T-1C DRAM) and allows on-demand routing of static and radio-frequency (rf) signals to individual devices. The architecture is composed of individual cells each containing a control field-effect transistor (\lq\lq control FET\rq\rq) and a quantum dot (QD) device. In our experiments, the QDs are themselves formed in the channel of a nanowire FET, integrated on the same chip as the control FETs and fabricated using the same CMOS processes. When not addressed, each cell can be used as a node to store charge on the QD device gate that allows trapping single-electrons in the QD device with a time constant approaching $1$~s. In this Article, we demonstrate random access and readout of two individual cells at cryogenic temperatures. For readout, we use capacitive gate-based radio-frequency reflectometry that turns the gate defining the quantum dot into a sensor by coupling it to a lumped-element resonant circuit~\cite{Colless2013,Gonzalez-Zalba2015,Ahmed2018Radio-FrequencySensing}. We characterize the bandwidth of the architecture in terms of a resonant frequency overlap of $13$~MHz, and find optimal operation voltage levels for the digital transistor. Moreover, we show dynamic readout of the cells and obtain charge stability maps sequentially. Finally, we propose an architecture for addressing large-scale device arrays with a quadratic reduction in the number of inputs.

%


\section{\label{sec:setup}Circuit characterization}

\begin{figure}
		\includegraphics[width=\linewidth]{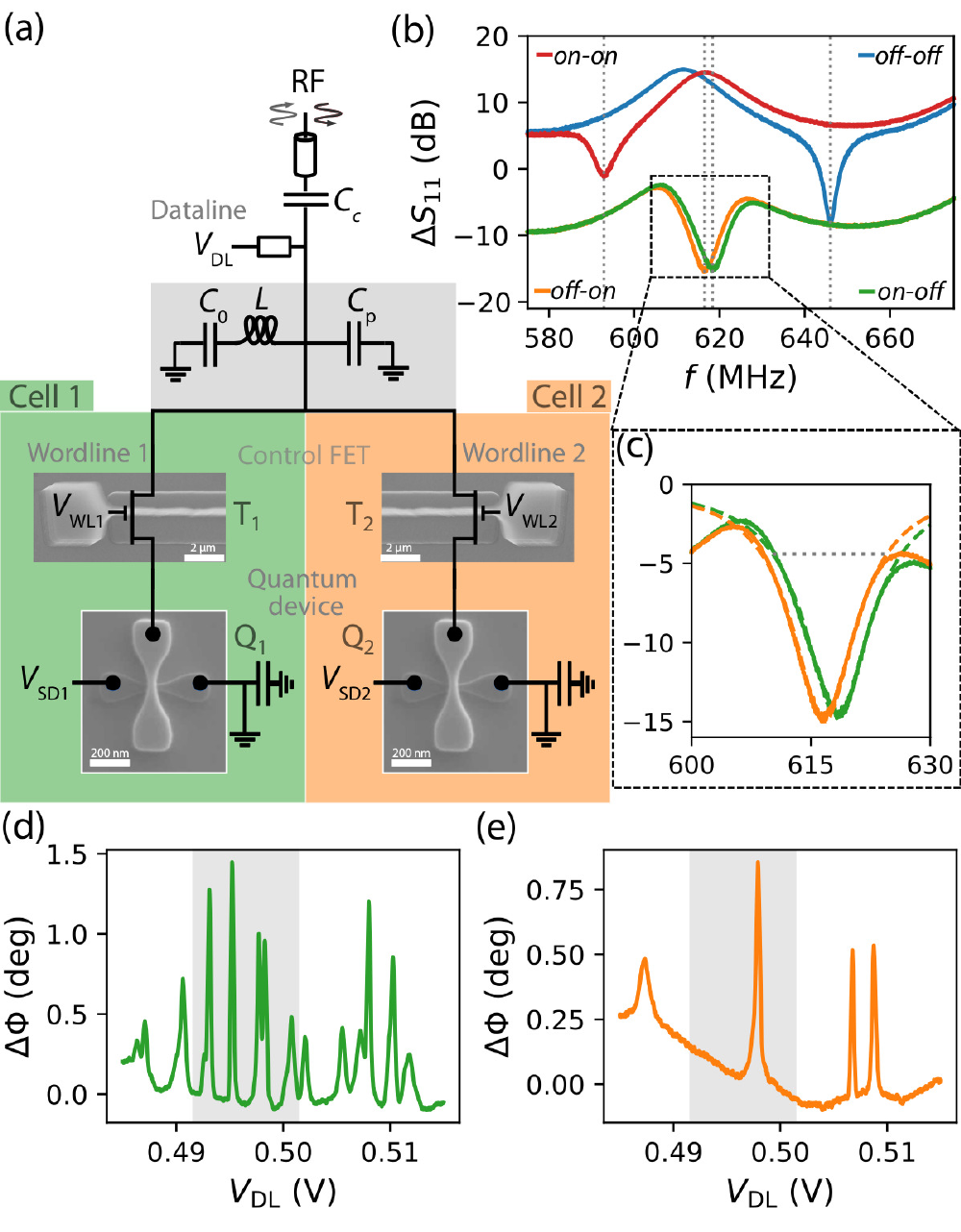}
        \caption{Setup and individual device characterization. \textbf{(a)} Sequential access circuit for gate-based radio-frequency (rf) readout. A single high frequency line and readout resonator (highlighted in grey) is connected to two cells (green and orange) consisting of one control FET, T$_i$, and quantum device, Q$_i$, per cell. T$_i$ enables selective readout of Q$_i$. \textbf{(b)} Reflection coefficient spectrum of the circuit for different control FET states (T$_1$ - T$_2$). Spectra for addressing a single cell have been shifted down by 15~dB for clarity. \textbf{(c)} Enlarged view of \textit{on-off} and \textit{off-on} state configurations with a spectral overlap and resonance fit indicated. \textbf{(d)} Phase response of Q$_1$ as a function of $V_\mathrm{DL}$ for $V_\mathrm{WL1}=1.2$~V and $V_\mathrm{WL2}=0$~V. \textbf{(e)} Phase response of Q$_2$ as a function of $V_\mathrm{DL}$ for $V_\mathrm{WL1}=0$~V and $V_\mathrm{WL2}=1.2$~V. The regions in grey highlight charge transitions we focus on in further measurements.}
        \label{fig:setup}
\end{figure}

We show the sequential access circuit in Fig.~1(a). It consists of two CMOS single-electron memory cells~\cite{Schaal2018} (cell 1(2) in green(orange)) connected to a single bias line and a lumped-element radio-frequency resonator for readout (resonator components are highlighted in grey). 
Each memory cell is made from two transistors which we refer to as Q$_i$ and T$_i$. Q$_i$ is a 60-nm-wide silicon nanowire transistor with a short channel length ($25$ and $30$~nm for cell 1 and 2 respectively). Such devices are routinely used to trap single-electrons in quantum dots that form at the top most corners of the nanowire channel when operated in the sub-threshold regime at cryogenic temperatures~\cite{Voisin2014,Gonzalez-Zalba2016}. Transistor T$_i$ is a wider device with a channel width of $10\, \mu$m and gate length of $25$~nm and $30$~nm for cell 1 and 2 respectively which we refer to as the control FET. The four transistors are manufactured using fully-depleted silicon-on-insulator (FD-SOI) technology following standard CMOS processes. They are located on the same chip and are connected via bond wires (see Methods for details of the fabrication). 

We label the primary inputs of the circuit as data and word lines in analogy with memory chips.
Each cell has one word line $V_{\mathrm{WL}i}$ which connects to the gate of the control FET T$_i$ allowing control over the channel resistance. The data line $V_\mathrm{DL}$ is shared among the two cells and allows control over the gate voltage on Q$_i$ conditional on the state of T$_i$. Additionally, a voltage applied to the silicon substrate, $V_\text{BG}$, acts as a back-gate. 
Switching T$_i$ to the \textit{on} state while keeping all the remaining T$_j$ \textit{off} allows for individual addressing of a single quantum device Q$_i$. Multiple devices can be addressed sequentially by timing the voltages on T$_i$ accordingly, as we demonstrate further below.

To read the quantum state of the devices sequentially, we connect a lumped-element $LC$ resonator in parallel with the memory cells and use radio-frequency reflectometry to probe the resonant state of the combined circuit~\cite{Schaal2018}. 
The natural frequency of the resonator $f_0$ is given by $f_0=1/2\pi\sqrt{LC_\mathrm{T}}$ where $C_\mathrm{T}$ is the total capacitance of the system that includes, in particular, the state-dependent quantum or tunneling capacitance of any quantum device which is connected to the \textit{LC} circuit~\cite{Mizuta2016} via the control FETs. We drive the resonator with a small ($-90$~dBm) radio-frequency signal at frequency $f_c$ that we coupled into the data line via a bias tee.  
The whole circuit is operated in a dilution refrigerator with a base temperature of $15$~mK.

Next, we show the frequency dependence of the circuit's reflection coefficient $S_{11}$ in Fig.~1(b)
for the four possible states of the two control FETs. A dip in the reflection coefficient occurs when we drive the resonator at its natural frequency of oscillation, which shifts towards lower frequency for each T$_i$ in the \textit{on} state due to the additional circuit capacitance introduced by the enabled cell (by approximately $28$~MHz, see Supplementary Table S1 for resonance parameters). 
Most importantly, we observe a large spectral overlap of $13$~MHz with $3$~dB bandwidth in the enlarged view in Fig.~1(c) when addressing one cell at a time. 
Spectral overlap is paramount to dynamical multi-qubit readout as it means that both cells can be read using the same input frequency, while the degree of overlap determines the bandwidth of the architecture.
In addition to the resonance frequency shift, we observe a reduction in the loaded quality factor $Q_\text{L}$ from a value of 96, when both T$_1$ or T$_2$ are in the \textit{off} state, to a value of 40, when either T$_1$ or T$_2$ are in the \textit{on} state. An \textit{on}-state $Q_\text{L}$ of 40 is comparable to previous experiments with~\cite{Schaal2018} and without~\cite{Gonzalez-Zalba2015} control circuit.

Based on the spectra shown in Fig.~1(c), we select a carrier frequency $f_c=615$~MHz to probe the state of the quantum devices. When using radio-frequency reflectometry, changes in the complex impedance of the circuit are probed by driving the circuit close to resonance while monitoring the phase and magnitude of the reflected signal (see Methods for details of the circuit). Changes in the capacitance of the quantum device $\Delta C$, attributed to tunneling of single electrons, are detected through changes in the reflected phase $\Delta \Phi = -2 Q \Delta C/C_\mathrm{T}$~\cite{Ahmed2018Radio-FrequencySensing}. 
In Fig.~1(d-e), we observe phase shift peaks as we change $V_\mathrm{DL}$ that corresponds to regions of charge instability in Q$_i$. At these voltages, single electrons cyclically tunnel between the QDs in the channel and the source or drain electron reservoirs in Q$_i$. For each measurement only one T$_i$ is set to the \textit{on} state while the other is \textit{off}. 
Below we perform further measurement focusing on a particular region of this stability diagram (highlighted in Fig.~1(d-e)) for both quantum devices.

\section{\label{sec:digital}Digital transistor operation parameters}

\begin{figure}
		\includegraphics[width=\linewidth]{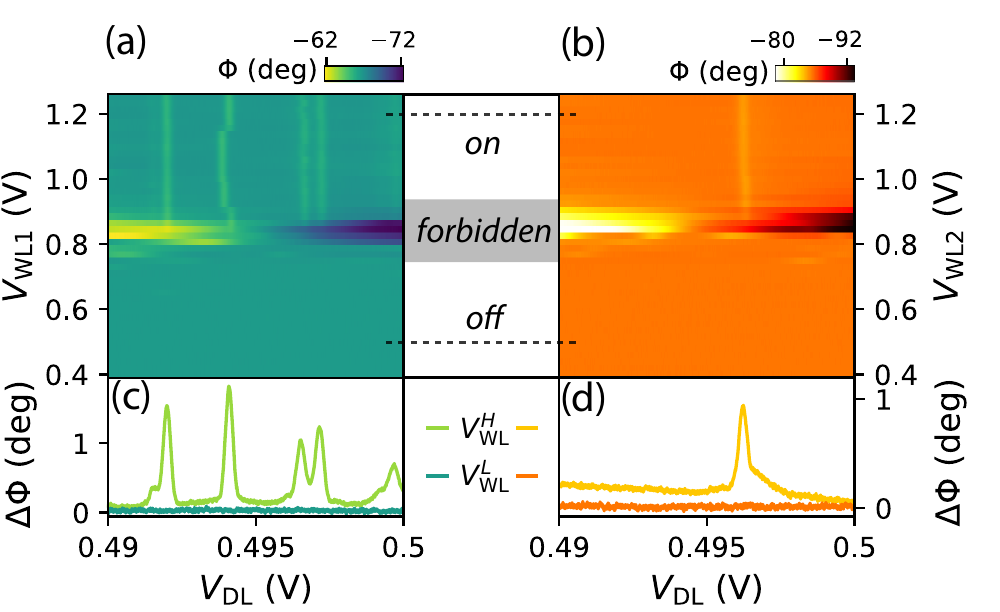}
        \caption{Control transistor logic states. \textbf{(a)} Phase response of Q$_1$ as a function of $V_\mathrm{WL1}$ and $V_\mathrm{DL}$ ($V_\mathrm{WL2}=0$~V). \textbf{(b)} Phase response of Q$_2$ as a function of $V_\mathrm{WL2}$ and $V_\mathrm{DL}$ ($V_\mathrm{WL1}=0$~V). For both cells we observe QD-to-reservoir transitions at large $V_\mathrm{WL}$ corresponding to the logical \textit{on} state of the digital transistor. A \textit{forbidden} region of large background signal is found upon approaching the control FET threshold voltage. A region of no signal below threshold corresponds to the \textit{off} state. \textbf{(c-d)} Line cuts at $V_\mathrm{WL}^\text{L}=0.5$~V and $V_\mathrm{WL}^\text{H}=1.2$~V, indicated by dashed lines in (a-b), that highlight the difference between the two digital states for each device.}
\end{figure}

In this section, we focus on the digital operation of the control transistors with the aim to find the optimal voltage levels. For a dynamical random-access readout scheme, T$_i$ should fulfill several requirements: In the \textit{on} state, T$_i$ should be sufficiently conductive to allow high-sensitivity gate-based readout of the selected quantum device. In the \textit{off} state, T$_i$ should be sufficiently resistive to block the rf signal to deselected cells and to retain the charge on Q$_i$'s gate for the duration operations are being performed in other cells.


As a first step towards dynamically operating the circuit, we identify suitable \textit{on} and \textit{off} state voltages for the control FET gate (i.e. the \textit{high}, $V_{\text{WL}i}^\text{H}$, and \textit{low}, $V_{\text{WL}i}^\text{L}$, signal levels). In Fig.~2(a-b), we show the phase of the reflected signal from the resonator as a function of $V_\mathrm{DL}$ and $V_{\mathrm{WL}i}$ in the $V_\mathrm{DL}$ areas highlighted in Fig.~1(d-e). We can identify three regions: The \textit{on} region for $V_{\mathrm{WL}i}> 0.9$~V, where we observe single electron tunneling, the \textit{off} region for $V_{\mathrm{WL}i} < 0.7$~V, where we observe no transitions and finally, for 0.7 V$<V_{\mathrm{WL}i}<0.9$~V the \textit{forbidden} region. In the latter, T$_i$ is in the saturation regime, where, due to the voltage-dependence gate capacitance of the control FET, the phase varies largely~\cite{Rossi2016}. This region should be avoided when assigning voltage levels. To highlight the different response of the resonator in the digital \textit{on} and \textit{off} states, we show the phase change $\Delta\phi$ as a function of $V_\text{DL}$ for cell 1 and 2 in Fig.~2(c-d), respectively, at $V_{\text{WL}i}^\text{L(H)}=0.5(1.2)$~V (as indicated by the dashed black lines in Fig. 2(a-b)).

We note the close similarity between the operation voltage levels of both T$_i$ for addressing the quantum devices Q$_i$ at millikelvin temperature. In a scaled up architecture, with increasing circuit complexity, reproducible electrical characteristics between cells will be essential.

\section{Dynamic Operation}

\subsection{\label{sec:memory}Charge retention time}

\begin{figure}
		\includegraphics[width=\linewidth]{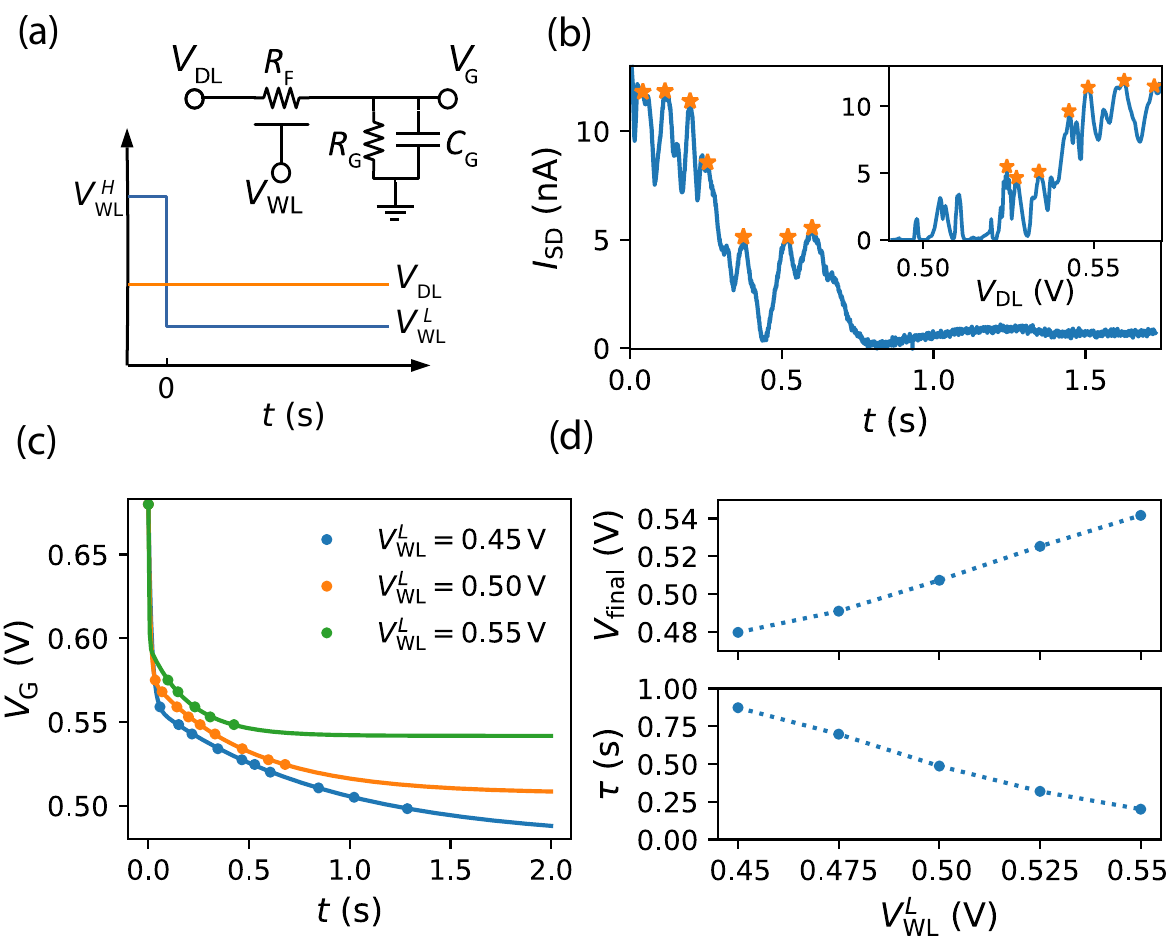}
        \caption{Charge retention analysis. \textbf{(a)} Equivalent circuit of a single memory cell and pulsing scheme for charge retention analysis. 
        \textbf{(b)} Source-drain current $I_\mathrm{SD}$ through the quantum device as a function of time after switching the control transistor to the \textit{off} state $V_\mathrm{WL}^\text{L}=0.5$~V ($V_\mathrm{SD}=2$~mV). Single electron transitions are observed and peak positions are indicated by stars. The inset shows $I_\mathrm{SD}$ as a function of $V_\mathrm{DL}$ where the same transitions are observed and indicated by stars.
	\textbf{(c)} Decay of the voltage on the quantum device gate $V_\text{G}$ as a function of time once the control transistor is switched off obtained from the peak positions (stars) in (b) (circles) with double exponential fit (solid line).
   		\textbf{(d)} Quasi-static gate voltage $V_\mathrm{final}$ and time constant $\tau$ as a function of $V_\mathrm{WL}^\text{L}$ obtained from fits (see Eq. \ref{eq:Vg}) to data in (c) (circles). Dashed lines are guides to the eye.}
\end{figure}

Random access of a single cell can be achieved by switching the selected T$_i$ \textit{on} while all other T$_j$ are \textit{off}. Since the data line voltage $V_\mathrm{DL}$ is shared among the cells, the gate voltage on all deselected Q$_j$ floats and decays over time while addressing cell $i$. Floating gate charge storage, which is also found in modern dynamic random access memory (DRAM) chips, is an important feature of dynamic readout and its associated charge retention time sets the maximum time to perform operations on other cells before the information is lost.

In this section, we characterize the discharge of one cell in order to determine an appropriate voltage refresh rate. We consider the equivalent circuit model of the memory cell as shown in Fig.~3(a) with a $V_\mathrm{WL}$-dependent channel resistance $R_\mathrm{F}$ of the control FET and a gate resistance $R_\text{G}$ of the quantum device combined with a capacitance $C_\text{G}$, realizing charge storage on the quantum device gate. $C_\text{G}$ has a contribution from the FET, the quantum device gate and the parasitic capacitance of the inter-connection. The latter dominates in this experiment (see Supplementary Table S1 and Note). In the \textit{on} state, the voltage $V_\text{G}$ on the gate capacitor is charged to $V_\text{DL}$, while in the \textit{off} state $V_\text{G}$ decays over time as

\begin{align}
	V_\text{G}(t)= V_\mathrm{final} \left[1+\frac{R_\mathrm{F}}{R_\text{G}} \exp \left(- \frac{t}{\tau} \right) \right].
	\label{eq:Vg}
\end{align}

\noindent Here $\tau=\frac{C_\text{G}R_\text{G}R_\text{F}}{R_\text{G}+R_\text{F}}$ is the circuit time constant and $V_\mathrm{final}=\frac{V_\mathrm{DL}R_\text{G}}{(R_\mathrm{F}+R_\text{G})}$ is the equilibrium voltage at the gate of the quantum device at $t\to\infty$. Since $\tau$ and $V_\mathrm{final}$ depend on $R_\mathrm{F}$, and thus on the operation voltage level $V_\mathrm{WL}$, we proceed by investigating their functional dependence further in order to find the optimal voltage operation point to maximize the charge retention time.

We monitor the discharge of the cell in a pulsed experiment by measuring the source-drain current $I_\mathrm{SD}$ through the quantum device over time. As shown in Fig.~3(a), we keep $V_\mathrm{DL}$ constant while $V_\mathrm{WL}$ switches from the \textit{high} level ($V_\mathrm{WL}^\text{H}$) to a \textit{low} level ($V_\mathrm{WL}^\text{L}$) at $t=0$. We set the pulse amplitude to 0.5~V and vary the pulse offset ensuring that the transistor remains \textit{on} in the \textit{high} part of the pulse. We show an example discharge measurement for $V_\mathrm{WL}^\text{L}=0.5$~V in Fig.~3(b), where several single electron transitions (indicated by stars) can bee observed in $I_\mathrm{SD}$ over time. After $1.5$~s the current settles to a value determined by $V_\mathrm{final}$. We compare the discharge data with a measurement of the same single electron transitions of the device as a function of $V_\mathrm{DL}$ in quasi-static conditions as shown in the inset of Fig.~3(b). By matching peaks in the decay over time to peaks as a function of $V_\mathrm{DL}$ we reproduce the dynamics of the voltage on the quantum device gate $V_\text{G}(t)$ as shown in Fig.~3(c) for multiple values of $V_\mathrm{WL}^\text{L}$. At $t=0$, we observe an initial fast decay, possibly due to cross-talk, followed by a slow decay characterized by Eq.~\ref{eq:Vg}. We fit a double exponential to capture the fast and slow dynamics (see Supplementary Note) and extract $V_\mathrm{final}$ and $\tau$ from the slow decay which we show in Fig.~3(d) as a function of $V_\text{WL}^\text{L}$. We observe that as $V_\mathrm{WL}^\text{L}$ increases ($R_\mathrm{F}$ decreases) $V_\mathrm{final}$ becomes larger due to the voltage divider characteristic of the cell (see panel (a)). In the case of $\tau$, we observe a reduction from $0.9$~s to $0.2$~s. Both trends are captured in our model. We note that the time constant could be increased further for a larger capacitance $C_\text{G}$.

While initially it may seem beneficial to select a low $V_\text{WL}^\text{L}$ level to maximize $\tau$, the retention or refresh time is determined by the maximum tolerable gate voltage drop of the cell, $\delta V$, which has to be accessed given a specific qubit implementation. Using our model and defining the voltage drop ratio $a=\delta V/V_\text{DL}$ and the resistance ratio $r=(R_\text{F}+R_\text{G})/R_\text{F}$, we find that the retention time is given by $t_\text{r}=R_\text{G}C_\text{G}\ln[(1-ar)^{-1}]/r$ which is a monotonically increasing function of $r$. Hence, $t_\text{r}$ is maximized by operating at large $V_\text{WL}^\text{L}$ while remaining in the \textit{off} regime. 

In this analysis, it is important to note that we keep $V_\text{DL}$ constant which is approximately what will happen when addressing multiple quantum devices with similar operating voltages. Such operation is a particular feature of our proposal and differs from the 1T-1C DRAM read protocol where $V_\mathrm{DL}$ is typically set to half the maximum voltage stored in the capacitor. Such voltage level maximizes the readout signal and the retention time of both the uncharged and charged memory state of the capacitor~\cite{DavidTaweiWang2005}. In our proposal, we operate exclusively at the charged state of Q$_i$ where retention time is maximized by selecting a less resistive FET, i.e. $V_\text{WL}^\text{L}$ as large as possible while remaining in the \textit{off} state.
For sequential readout, as demonstrated further below, we select $V_\mathrm{WL}^\text{L} = 0.5$~V to enhance the retention time in the \textit{off} state while preserving good noise margins.

\subsection{\label{sec:power}Dynamic power dissipation}

Having extracted the digital voltage levels for the control FETs, we can estimate the dynamic power dissipation of the architecture. This parameter is important when operating at the base temperature of a dilution refrigerator where limited cooling power, typically 400~$\mu$W, is available. The dynamic power when switching T$_i$ is $P=C_\mathrm{cell} f_\mathrm{op}\Delta V^2$, where $\Delta V= V_\mathrm{WL}^\text{H}-V_\mathrm{WL}^\text{L}$, $C_\text{cell}$ is the additional capacitance introduced by switching T$_i$ to the \textit{on} state and $f_\mathrm{op}$ is the switching frequency. The maximum $f_\mathrm{op}$ is given by the bandwidth of the architecture and $C_\text{cell}$ by the gate area of the FET plus the parasitic capacitance of the interconnect between T$_i$ with Q$_i$. Given our selection of $\Delta V=0.7$~V, the bandwidth 13~MHz and $C_\text{cell}=70$~fF (see Supplementary Table S1), we estimate 0.4~$\mu$W/cell which will allow operating 1,000 cells simultaneously. The power dissipation can be reduced by pushing the voltage levels closer to the \textit{forbidden} region $\Delta V=0.4$~V, thus reducing the noise level margins, and by minimizing $C_\text{cell}$ by optimizing the circuit layout. The minimum $C_\text{cell}$ will be the dominated by the geometric capacitance of T$_i$ of 8~fF. At the same frequency of operation the dissipation will be $P\approx 0.05$~$\mu$W/cell extending the number of cells that can be operated at the same time to 20,000. When reducing the circuit capacitance the noise associated with charging the cell via the control FET resistance should be considered $V_\text{RMS}=\sqrt{k_B T/C_\text{cell}}$. In this experiment we estimate the noise at $1.7\, \mu$V (15~mK base temperature).

\subsection{\label{sec:Multiplex}Sequential readout}

\begin{figure}
		\includegraphics[width=\linewidth]{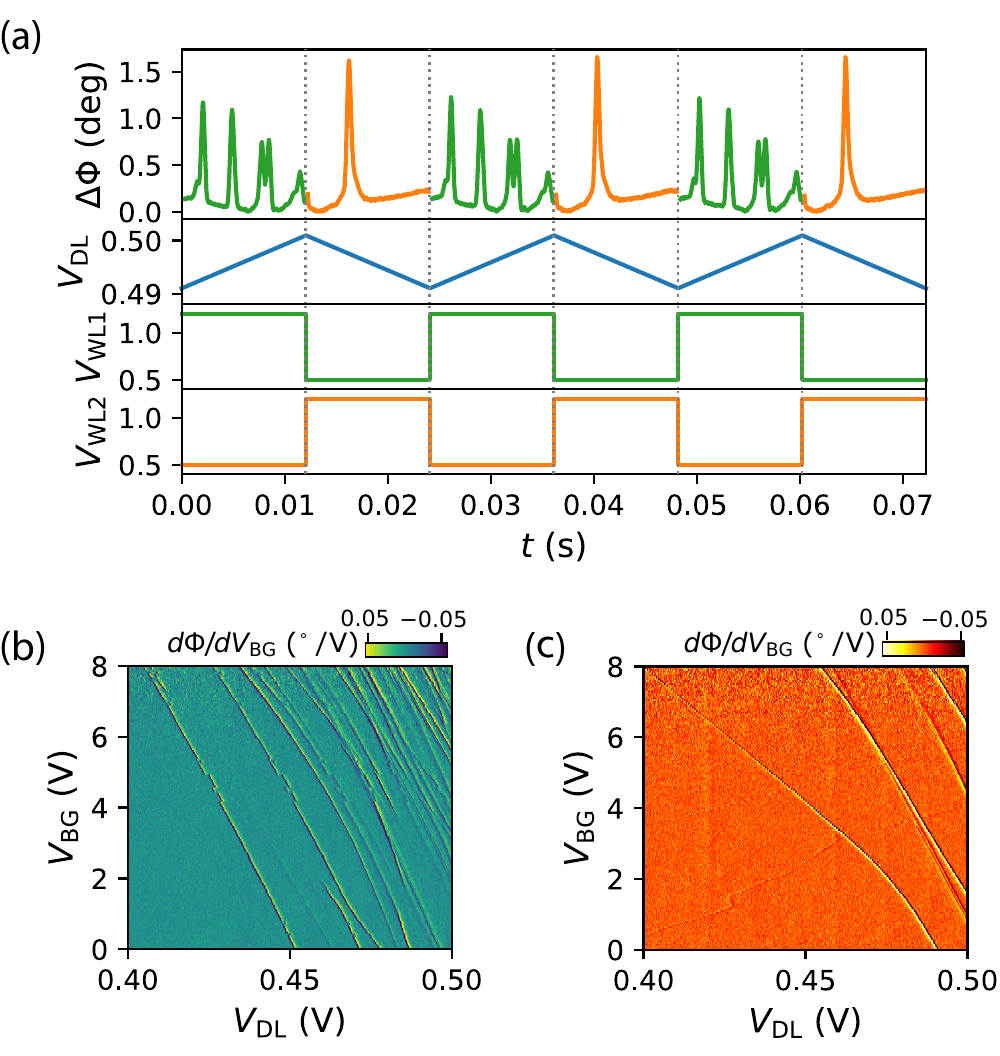}
        \caption{Dynamic readout. \textbf{(a)} Pulse scheme for sequential readout and phase response of Q$_1$ and Q$_2$. $V_\mathrm{DL}$ is ramped up and down while $V_\mathrm{WL1}$ and $V_\mathrm{WL2}$ are alternating between \textit{high} and \textit{low} states. Pulses are synchronized such that QD reservoir transitions from Q$_1$ are obtained when $V_\mathrm{DL}$ is ramped up while Q$_2$ is measured when $V_\mathrm{DL}$ is ramped down. \textbf{(b)-(c)} Differential phase response obtained sequentially from both cells as a function of data line and back-gate voltages. 
        }
\end{figure}

We now turn to demonstrate sequential dynamic readout of quantum devices in two memory cells. 
We show the pulsing scheme to dynamically read both memory cells in Fig.~4(a). 

In the first half of the cycle, from 0 to 12~ms, we set T$_1$ and T$_2$ to the digital \textit{on} and \textit{off} states respectively. Simultaneously, we apply an analogue signal to the common data line $V_\mathrm{DL}$ (blue trace) that ramps up the gate voltage on the data line (now connected to Q$_1$). We read the signal dispersively using gate-based readout and detect peaks in the phase due to single-electron transitions between a QD and a reservoir in Q$_1$. 

In the second half of the cycle, from 12 to 24~ms, we invert the digital voltages on T$_1$ and T$_2$ such that we can now detect the transitions in Q$_2$ as we ramp down the analogue signal on the data line. 
The QD-to-reservoir transitions in the phase response are identical to those measured in a static experiment shown in Fig.~1(d-e). 
The rf modulation frequency and amplitude is kept constant throughout the measurement. There is a phase offset between the signal detected from Q$_1$ and Q$_2$ due to a small difference in reflection coefficient between cells (see Fig.~1(c)). We therefore show the phase difference $\Delta \Phi$ in Fig.~4(a) (see Supplementary Figure S2).

Using this interleaved pulsing scheme for sweeping $V_\mathrm{DL}$ combined with additional stepping of $V_\mathrm{BG}$ after each cycle, we obtain the charge stability map of both Q$_1$ and Q$_2$ sequentially as shown in Fig.~4(b-c). The transitions observed in the measurement suggest formation of multiple QDs in both cells (see Supplementary Figure S1 for additional scans), i.e. corner dots~\cite{Voisin2014,Betz2014}.
Due to filtering of the lines delivering the control FET signals $V_\mathrm{WL1,2}$ and data line signal $V_\mathrm{DL}$, $f_\text{op}$ was limited to $1$~kHz in this experiment (see Supplementary Figure S2).

\section{\label{sec:Large}Discussion: Scaling up the architecture}

In this Article, we have presented a CMOS dynamic-random-access architecture for radio-frequency readout of quantum dot devices at millikelvin temperature. In particular, we performed sequential dispersive readout of individual quantum devices in a two-cell layout. In this section, we describe how a larger ($N \times M$) array of randomly addressable single-electron memory cells could be constructed, making use of the sequential gated readout we demonstrate above, combined with frequency multiplexing. As shown in Fig. 5, the two dimensional array is distributed in rows ($i$) and columns ($j$). A specific qubit in row $i$ and column $j$, Q$_{ij}$, can be addressed by a word line $V_{\text{WL}j}$, that controls the voltage on the gate of transistor T$_{ij}$, and by a data line voltage $V_{\text{DL}i}$, that controls the gate potential on the qubit.
Additionally, each row is connected to a different $LC$ resonator for readout. Distinct resonant frequencies $f_i$ are achieved using different values for the inductance of each resonator $L_i$. Note that the inductors have a low high-frequency impedance to ground in parallel with a parasitic capacitor. Each resonator can be probed simultaneously using frequency multiplexing techniques~\cite{Hornibrook2013} such that the whole array only requires a sole high-frequency line for readout.
The array can be operated dynamically in a random access manner. When a specific word line $j'$ is activated qubits Q$_{ij'}$ can be read simultaneously for all $i$ or their gate voltages are refreshed using the specific data line voltages $V_{\mathrm{DL}i}$. 
Each qubit may be fabricated such that it is in close proximity to other qubits to perform two-qubit operations~\cite{Zajac2017b,Watson2018}. This possibility is indicated by the circular connection at the source of each quantum device. The array particularly suits one dimensional chains of interacting quantum devices such as they are distributed along the rows of the array~\cite{Betz2016, Zajac2016, Jones2018}.

\begin{figure}
	\centering
	\includegraphics{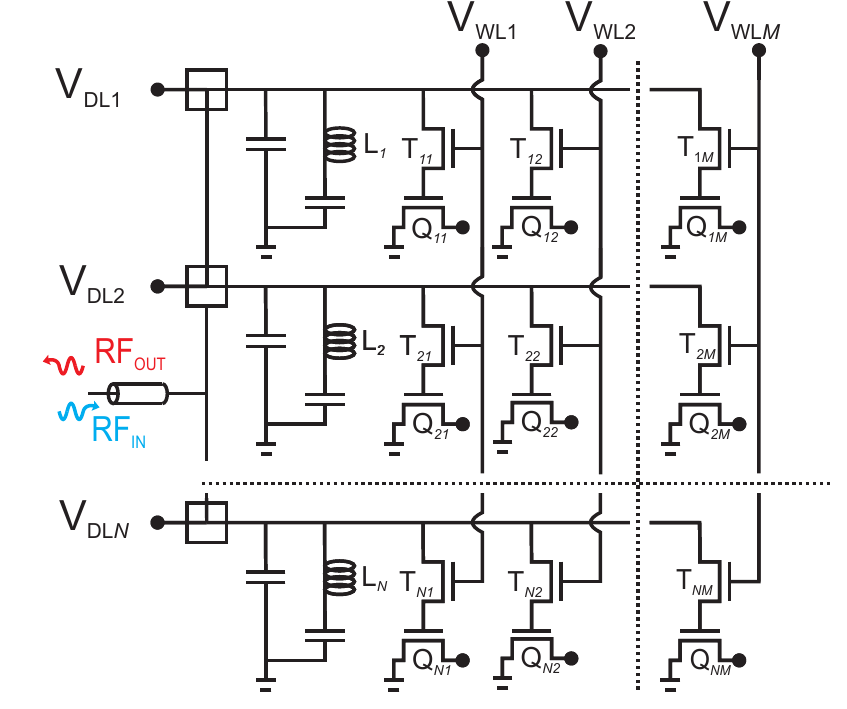}
	\caption{A $N \times M$ 2D array of single electron memory cells for sequential multi-qubit readout. Control transistors T$_{ij}$ can be controlled by word line voltages $V_{\text{WL}j}$. Qubits Q$_{ij}$ can be controlled with data lines voltages $V_{\text{DL}i}$. A single high-frequency line allows simultaneous readout via frequency-multiplexed radio-frequency reflectometry. The array can be extended in number of rows and columns limited by the spectral overlap of the data-line resonators.} 
	\label{fig5}
\end{figure}

For further optimization of the architecture a more detailed understanding of each circuit component at low temperature is necessary, including cross-talk mitigation and a circuit equivalent model of both the control FETs and the quantum devices at the desired operation bias and frequency. Such models could be used in combination with ECAD mixed-circuit design tools to produce an integrated circuit version of the 2D array. Moreover, we note when applying this proposed architecture to particular qubit implementations, where small voltage drifts may induce undesired effects such as unwanted decoherence, further optimization could be necessary to find an optimal trade off between voltage noise, retention time and power dissipation depending on the cell's capacitance.
Finally, we note that the signal-to-noise of gate-based readout -- limited by the quality factor of the resonant circuit -- could be improved by using superconducting spiral inductors~\cite{Ahmed2018Radio-FrequencySensing}. There is a potential for such inductors to be made CMOS compatible using TiN allowing on-chip integration. The footprint of such inductors can be reduced when operating at higher frequencies which could ultimately enable an integrated solution for sensitive readout of large-density qubit arrays.

\section*{\label{sec:Methods}Methods}

\subsection*{Fabrication details}
All CMOS transistors used in this study were fabricated on SOI substrates with a $145$-nm-thick buried oxide and $10$-nm-thick silicon layer. The silicon layer is patterned to create the channel by means of optical lithography, followed by a resist trimming process. All transistors share the same gate stack consisting of $1.9$~nm HfSiON capped by $5$~nm TiN and $50$~nm polycrystalline silicon leading to a total equivalent oxide thickness of $1.3$~nm. The Si thickness under the HfSiON/TiN gate is $11$~nm. After gate etching a SiN layer ($10$~nm) was deposited and etched to form a first spacer on the sidewalls of the gate. 18-nm-thick Si raised source and drain contacts were selectively grown before the source/drain extension implantation and activation annealing. A second spacer was formed followed by source/drain implantations, activation spike anneal and salicidation (NiPtSi). The wide channel control FETs $T_i$ and nanowire quantum devices $Q_i$ are connected via on-chip aluminium bond wires.

\subsection*{Measurement setup}
Measurements were performed at base temperature of a dilution refrigerator ($15$~mK). Low frequency signals ($V_{\mathrm{SD}}$, $V_{\mathrm{DL}}$, $V_{\mathrm{WL1,2}}$) were delivered through filtered cryogenic loom while a radio-frequency signal for gate-based readout was delivered through an attenuated and filtered coaxial line which connects to a on-PCB bias tee combining the rf modulation with $V_{\mathrm{DL}}$. The resonator is formed from a  $82$~nH inductor and the sample's parasitic capacitance to ground in parallel with the device. The inductor consists of a surface mount wirewound ceramic core (EPCOS B82498B series) and the PCB is made from $0.8$~mm thick RO4003C with immersion silver finish.
The reflected rf signal is amplified at $4$~K (LNF-LNC0.6\_2A) and room temperature followed by quadrature demodulation (Polyphase Microwave AD0540B) from which the amplitude and phase of the reflected signal is obtained (homodyne detection).

	\begin{acknowledgments}
	We would like to thank Sebastian Pauka for helpful discussions. This research has received funding from the European Union's Horizon 2020 research and innovation programme under grant agreement No 688539 (http://mos-quito.eu) and Seventh Framework Programme (FP7/2007-2013) through Grant Agreement No. 318397 (http://www.tolop.eu.); as well as by the Engineering and Physical Sciences Research Council (EPSRC) through the Centre for Doctoral Training in Delivering Quantum Technologies (EP/L015242/1) and UNDEDD (EP/K025945/1). M.F.G.Z. and A.R. acknowledge support from the Winton Programme for the Physics of Sustainability and Hughes Hall, University of Cambridge.
\end{acknowledgments}

\section*{Author contributions}
S.S. and M.F.G.-Z. devised the experiment. S.S., A.R. and M.F.G.-Z. performed the experiments; S.B. fabricated the sample; S.S. did the analysis and prepared the manuscript with contributions from A.R., J.J.L.M. and M.F.G.-Z.

\section*{Data availability}
The data that support the findings of this study are available from the corresponding author upon reasonable request.


%

\end{document}


	
	
	
	\title{A CMOS dynamic random access architecture for radio-frequency readout of quantum devices:\\
	Supplementary Material}
	
	\author{S. Schaal}
	\email{simon.schaal.15@ucl.ac.uk}
	\affiliation{London Centre for Nanotechnology, University College London, London WC1H 0AH, United Kingdom}
	\author{A. Rossi}
	\affiliation{Cavendish Laboratory, University of Cambridge, J. J. Thomson Ave., Cambridge, CB3 0HE, United Kingdom}
	\author{S. Barraud}
	\affiliation{CEA, LETI, Minatec Campus, F-38054 Grenoble, France}
	\author{J. J. L. Morton}
	\affiliation{London Centre for Nanotechnology, University College London, London WC1H 0AH, United Kingdom}
	\affiliation
	{Department of Electronic \& Electrical Engineering, University College London, London WC1E 7JE, United Kingdom}
	\author{M. F. Gonzalez-Zalba}
	\email{mg507@cam.ac.uk}
	\affiliation
	{Hitachi Cambridge Laboratory, J.J. Thomson Avenue, Cambridge CB3 0HE, United Kingdom}
	
\maketitle

	\section*{Note: Readout resonance parameters}
	
	We present the resonant frequency $f_0$, the frequency shift $\Delta f$, the total capacitance $C_\mathrm{T}$ and the loaded Q factor $Q_\mathrm{L}$ in Tab.~S1 extracted from Fig.~1(b) of the main text for each control FET configuration. 
	
	\begin{table}[ht]
		\centering 
		\begin{tabular}{c c c c c} 
			\hline
			Case & $f_\mathrm{0}$~(MHz) & $\Delta f$~(MHz) & $Q_\mathrm{L}$ & $C_\mathrm{T}$~(pF) \\ [0.5ex] 
			\hline 
			low-low & 645.9 & 0 & 96 & 0.740 \\ 
			high-low & 618.3 & 27.3 & 39 & 0.808 \\
			low-high & 616.7 & 29.2 & 40 & 0.812 \\
			high-high & 593.0 & 52.9 & 28 & 0.878 \\ [1ex] 
			\hline 
		\end{tabular}
		\label{table:resonances} 
		\caption{Resonance parameters extracted from Fig.~1(b) of the main text. $f_\mathrm{0}$ is the center frequency, $\Delta f$ is the resonant frequency shift, $Q_\mathrm{L}$ the loaded quality factor and $C_\mathrm{T}$ the total circuit capacitance obtained using a nominal inductance of $82$~nH.} 
	\end{table}
	
	By comparison with the geometric capacitance of the control FET of $8$~fF ($30$~nm $\times$ $10\, \mu$m, $1.3$~nm equivalent oxide) we find that most of the additional capacitance causing the frequency shift could be attributed to wire bonds between T$_i$ and Q$_i$. This indicates that a reduction of shifts of the resonance frequency could be reduced in an optimized circuit for improved resonant frequency matching and overlap.

	\section*{Note: Sequential measurement of Coulomb diamonds}

	\begin{figure}[ht!]
		\includegraphics[width=0.8\linewidth]{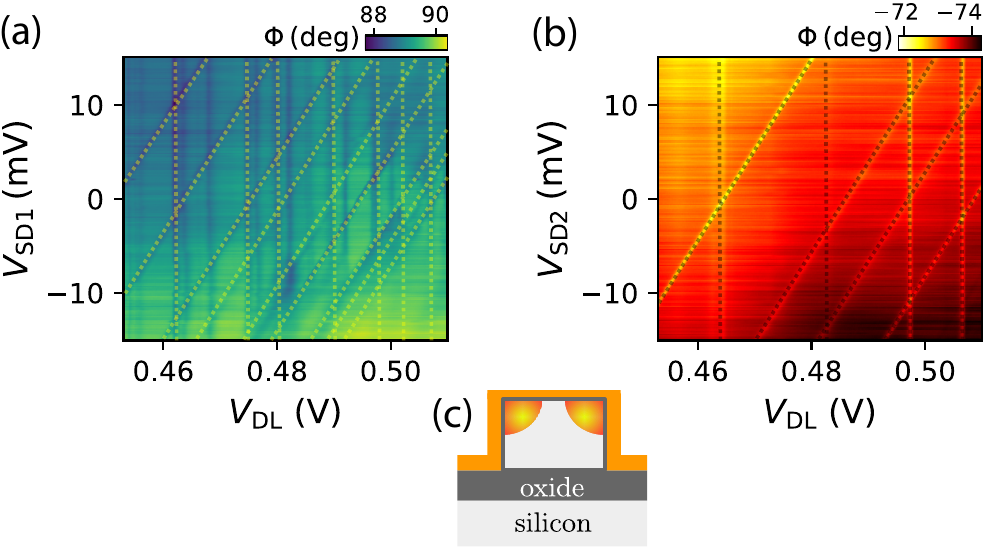}
		\caption{Dynamic readout. \textbf{(a)-(b)} Additional measurement as shown in Fig.~4 of the main text. Phase response obtained sequentially from both cells as a function of source-drain and topgate voltages.
			Coulomb diamonds are observed and highlighted using dashed lines. \textbf{(c)} Side view of the quantum device along the topgate showing formation of corner quantum dots.}
		\label{fig:coulomb}
	\end{figure}

	In addition to the sequential measurement as a function of top-gate and back-gate as shown in Fig.~4 of the main text we here present a sequential measurement of both cells as a function of top-gate and source-drain voltages. In Fig.~\ref{fig:coulomb}(a-b) we observe Coulomb diamonds compatible with formation of multiple quantum dots in both quantum devices. In these devices, quantum dots form in the corners of the silicon nanowire as indicated in the side-view along the top-gate in Fig.~\ref{fig:coulomb}(c). 
	
	\section*{Note: Filtering limitations}

	\begin{figure}[ht!]
	\includegraphics[width=0.7\linewidth]{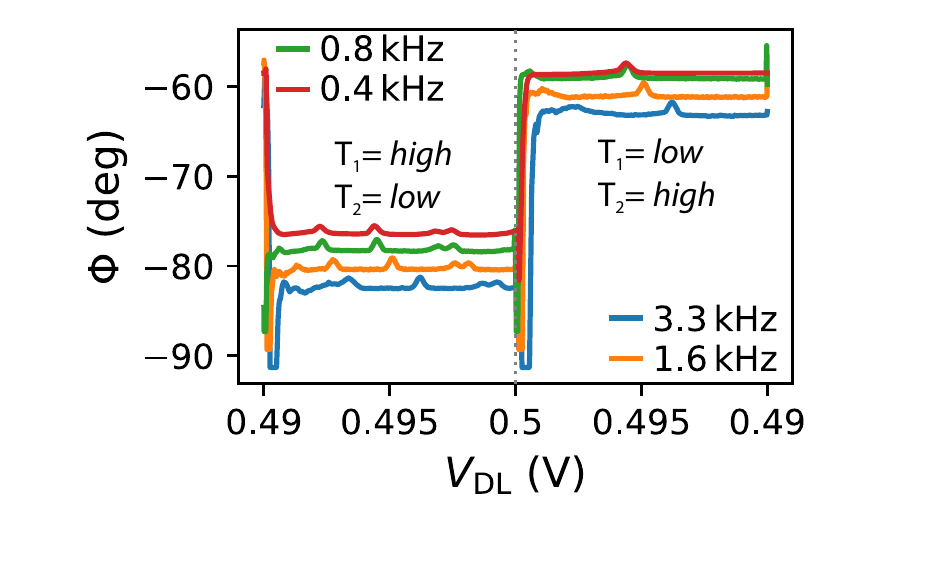}
	\caption{Filter limit. Phase response obtained in a dynamic readout experiment as shown in Fig.~4 of the main text for increasing control transistor switching frequency. There is a significant rise time of the pulses above $1$~kHz due to low-pass filtering of the lines. Traces are offset for clarification.}
	\label{fig:filter}
	\end{figure}

	In this section we analyze the impact of low-pass filtering of lines delivering the control FET switching voltages $V_{\mathrm{WL}i}$ and data-line voltage $V_\mathrm{DL}$.
	Fig.~\ref{fig:filter} shows a sequential measurement as shown in Fig.~4 of the main text where $V_\mathrm{DL}$ is ramped up and down while $V_\mathrm{WL1}$ and $V_\mathrm{WL2}$ are alternating between \textit{on} and \textit{off} states at the same frequency. Multiple traces represent measurement for increasing frequency and are offset for clarification. For measurements above $1$~kHz we observe a significant impact of the filtering on the rise time of the pulses representing the frequency limit for sequential readout in this experiment.
	By using coaxial lines for delivering $V_{\mathrm{WL}i}$ and $V_\mathrm{DL}$ the frequency limit could be improved up to a limit where fast switching would cause significant heating of the device.
	
	\section*{Note: Discharge of memory cell \& fitting procedure}
	

	In this section we explain the fitting procedure used to fit the decay of $V_G(t)$ as a function of time as shown in Fig.~3(b) of the main text.
	We observe an initial fast decay from $V_G(0)=0.68$~V followed by a slow decay with a quasi static final gate voltage once the control FET gate voltage is switched from $V_\mathrm{WL}^H$ ($1.2$~V) to $V_\mathrm{WL}^L$.
	We attribute the slow decay to discharge via the control FET ($R_F$) and gate dielectric ($R_G$) (see equivalent circuit in Fig.~3(a) of the main text).
	The fitting function consist of Eq.~1 of the main text parameterizing the slow decay in addition to an initial fast decay 
	\begin{align*}
		V_G(t) &=V_\mathrm{final} \left[1+\frac{R_\mathrm{F}}{R_G} \exp \left(- \frac{t}{\tau} \right) \right] + V_\mathrm{fast} \exp \left(-\frac{t}{\tau_\mathrm{fast}}\right)
	\end{align*} 
	with $\tau=\frac{C_G R_\mathrm{F} R_G}{R_\mathrm{F}+R_G}$ being the circuit's time constant and $V_\mathrm{final}=\frac{V_\mathrm{DL}}{(R_\mathrm{F}+R_G)/R_G}$ the final quasi static voltage.
	
The results for $V_\mathrm{final}$ and $\tau$ are shown in Fig.~3(e) of the main text and follow the expected trends.
The fast decay has a time constant of one milli-second and an amplitude of $\approx 0.1$~V which we attribute to cross-talk. To minimize the effect of the cross-talk on the analysis of the slow decay the pulse amplitude in the measurement presented in Fig.~3 of the main text is kept constant and the offset is varied to produce different \textit{low} levels while keeping the \textit{high} level above threshold. 
Cross-talk in the lines, PCB as well as the device could produce the fast initial decay. Charge injection when closing the control FET could play a role at the device level. We estimate the effect of charge injection by calculating the amount of charge under the control FET gate which is escaping onto the quantum device gate when the FET is turned off $\Delta V = \frac{Q_\text{channel}}{2 C_G}= \frac{C_\text{FET} (V_\text{WL}^\text{H} -V_\text{DL}-V_{th})}{2 C_G}\approx 0.02\, $V where $C_\text{FET}$ is the geometric capacitance of the FET and $C_G$ is the inter-connection and gate capacitance. Here we assume that the charge under the gate escapes equally to the source and drain which can be different in a real device. The voltage jump due to charge injection is reduced for smaller $C_\text{FET}$ and larger $C_G$ and charge injection can be mitigated using one NMOS and one PMOS control FET in parallel where the geometry of both can be optimized such that the negative voltage jump introduced when switching off the NMOS FET is canceled by the positive voltage change due to switching of the PMOS FET. Furthermore, cross-talk can be reduced by using coaxial lines to deliver control FET signals and careful PCB and chip design.